\documentclass{IAU-TrA}
\usepackage{graphicx}

\title[Solids and Their Surfaces] 
{}

\author[DIVISION~XII / COMMISSION~14 / WORKING GROUP] 
{}

\pubyear{2009}
\volume{Volume XXVIIA}
\pagerange{001--008}
\jname{Transactions IAU, Volume XXVIIA}
\editors{Karel A. van der Hucht, ed.}
\begin{document}

\maketitle

{\bf

\large
\noindent
DIVISION XII / COMMISSION~14 / WORKING GROUP                            \\ 
SOLIDS AND THEIR  SURFACES                                       \\

\normalsize

\begin{tabbing}
\hspace*{45mm}  \=                                                   \kill
CHAIR           \> Gianfranco Vidali                                  \\

\end{tabbing}

\vspace{3mm}

\noindent
TRIENNIAL REPORT 2009-2011
}

\firstsection 

\section{Introduction}

The {\it ISO} and {\it Spitzer} space observatories yielded a treasure trove of data on dust and ices covering dust grains. Now {\it Herschel}, and soon {\it SOFIA} and {\it ALMA}, will provide unprecedented views of the molecular world of the interstellar medium (ISM). It is on dust grains that key ISM molecules, such as  hydrogen, formaldehyde, methanol, and water  are formed. As a result of these new observations, there is a great need to know more about the interaction processes of atoms and molecules with dust grains. (The Proceedings of the 2010 NASA Laboratory Astrophysics Workshop 

\noindent(www-cfadc.phy.ornl.gov/nasa\verb+_+law/) give a good view of recent accomplishments in the study of atom/molecule - solid interactions as well as other aspects of laboratory astrophysics.)

In the last decade and a half there has been a tremendous increase of interest in laboratory studies of ISM processes occurring on interstellar dust grains. 
This has prompted the entrance into this field of a number of laboratories with a tradition in surface science. Besides the standard  probes that have been used in the past, techniques are now available that can give precise information at the atomic/molecular level about the formation of molecules on dust, including: Thermal Programmed Desorption (TPD),  Reflection Absorption Infrared Spectrometry (RAIRS), Resonant Enhanced Multiphoton Ionization (REMPI), and Atom Force Microscopy (AFM). These techniques yield information about the kinetics and energetics of atomic/molecular diffusion on and desorption from surfaces, the products of reactions, the ro-vibrational state of ejected products, and the morphology of the solid surfaces, respectively.  The success of research in atom/molecule/charged particle/photon-dust interaction has produced a surge of publications. Studies of interest to astrochemistry are now regularly published in chemical physics/ surface science journals.  A representative sample of such literature is listed below. It can be of use to astronomers and astrochemists in understanding the crucial steps of reactions on dust analogues. 

In theoretical research, there are two developments of note: the use of new stochastic tools to predict the molecule formation process on grains of different sizes, and the study of reaction mechanisms (Langmuir-Hinshelwood, Ealy-Rideal, and hot-atom) on surfaces of materials of astrophysical interest. Most of these studies pertain to hydrogen interaction with graphite/graphene/PAHs and appear in the chemical-physics literature.

\smallskip
Several  research groups that are currently working in this area are listed here, each with its group leader and main research focus:
\begin{itemize}

\item Catania Observatory, E. Palumbo (ions in ices)
\item Heriot-Watt University Edinburgh, M. R. S. McCoustra (desorption of mixed ices)
\item Hokkaido University, N. Watanabe / A. Kouchi (ice formation, photon-ice interaction)
\item Jet Propulsion Laboratory, A. Chutjian (ions on ices) 
\item Leiden University, H. Linnartz /  E. van Dishoeck (photodesorption from ices, water formation on ices)
\item Max Planck Institute for Astronomy, T. Henning (solids) 
\item NASA Ames Research Center, L. Allamandola (UV on ices, PAHs)
\item NASA Ames Research Center, F. Salama (dust exposure, dust formation, PAHs)
\item NASA - Goddard Space Flight Center, M. Moore (ions in ices)
\item Syracuse University USA, G. Vidali  (formation of $H_2$ and water on dust grain analogues)
\item University College London, W. A. Brown (desorption of mixed ices)
\item University College London, S. Price  (H$_2$ formation on graphite)
\item University of Cergy-Pontoise, J.-L. Lemaire (D$_2$ on ices, silicates)
\item University of Hawai'i, R. Kaiser (keV electron in ices)
\item University of Missouri, A. Speck (dust)
\item University of Virginia, R. Baragiola (ions in ices)

\end{itemize}

\vskip 0.1 true in
\section{Meetings}

\vskip 0.1 true in
During the reporting period, a number of meetings containing sessions about atomic and molecular interaction with surfaces have been held. They are often featured at regularly scheduled COSPAR, American Astronomical Society and Lunar and Planetary Institute meetings.  Unfortunately, meeting Web sites may no longer be accessible.

Important meetings (listed in inverse chronological order):
\begin{itemize}
\item European Conference on Laboratory Astrophysics, Paris, France, 2011
\item The Molecular Universe, IAU Symposium 280, Toledo, Spain 2011
\item Fifth Workshop on Titan Chemistry, Kauai, Hawai'i, 2011 

(http://www.chem.hawaii.edu/Bil301/Titan2011.html)
\item Herschel and the Characteristics of Dust in Galaxies, Lorentz Center, Leiden, Netherlands, 2011
\item Pacifichem, Honolulu, Hawai'i, USA, 2010
\item NASA Laboratory Astrophysics Workshop, Gatlinburg, TN, USA, 2010
\item Stormy Cosmos: the Evolving ISM from Spitzer to Herschel and Beyond, Pasadena, CA, USA, 2010 
\item WittFest: Origin and Evolution of Dust, Toledo, OH, USA, 2010
\item Molecules in Galaxies, Oxford Physics Conference Series, Oxford, United Kingdom, 2010
\item Dust and Ice: Their Roles in Astrophysical Environments, Univ. of Georgia, Athens, GA, USA, 2010
\item Recent Advances in Experimental and Observational Astrochemistry, Amer.Chem.Soc. Symposium, San Francisco, CA, USA, 2010
\item Infrared Emission, ISM and Star Formation, MPI, Heidelberg, Germany, 2010 
\item International Conference on Laboratory Astrophysics, Dunhuang, Gansu, China, 2009
\item Bridging Laboratory and Astrophysics: Molecules, Dust and Ices in Regions of Stellar and Planetary Formation, AAS 214$^{th}$, Pasadena, CA, 2009
\item The Chemical Enrichment of the Intergalactic Medium, Lorentz Center, Leiden, the Netherlands, 2009
\item Interstellar Surfaces: from Laboratory to Models, Lorentz Center, Leiden, the Netherlands, 2009
\end{itemize}

\section{Notable publications}
\vskip 0.1 true in
 Most of the works cited below regard the laboratory experiments and theories of photon and particle interaction with solid surfaces that are relevant to understanding similar processes occurring in space. Included in this selection are papers  about PAHs (Polycyclic Aromatic Hydrocarbons) that are relevant to atom/surface interactions. Additional information on PAHs can be obtained in the report of the Commission 14 Working Group on Molecular Data. Key observations that are related to dust are included. 
 
\vspace{3mm}
 
{\hfill Gianfranco Vidali}

{\hfill {\it Chair of Working Group}}

{ \hfill {\it Solids and Their Surfaces}}

\end{document}